\documentclass[prl,twocolumn,amsmath,amssymb,showpacs,superscriptaddress]{revtex4-2}
\usepackage{graphicx}
\usepackage{epstopdf}
\usepackage{dcolumn}
\usepackage{bm}
\usepackage[FIGTOPCAP]{subfigure}
\usepackage[usenames]{color}
\usepackage{mathtools}
\usepackage[all]{xy}

\begin{document}
	
	\title{Target Searches of Interacting Brownian Particles}
	
	\author{Sunghan Ro}
	\address{Department of Physics, Technion-Israel Institute of Technology, Haifa 3200003, Israel}
	\author{Juyeon Yi}
	\address{Department of Physics, Pusan National University, Busan 46241, Korea}
	\author{Yong Woon Kim} 
	\address{Department of Physics, Korea Advanced Institute of Science and Technology, Deajeon 34141, Korea}
	\address{Department of Physics, Massachusetts Institute of Technology, Cambridge, Massachusetts 02139, USA}
	
	
	\begin{abstract}
	We study the target search of interacting Brownian particles in a finite domain, focusing on the effect of inter-particle interactions on the search time. We derive the integral equation for the mean first-passage time and acquire its solution as a series expansion in the orders of the Mayer function. For dilute systems relevant to most target search problems, we analytically obtain the leading order correction to the search time and prove a universal relation given by the particle density and the second virial coefficient. Finally, we validate our theoretical prediction by Langevin dynamics simulations for the various types of the interaction potential. 
	\end{abstract}
	
	\maketitle
	
	Target search by random walkers is a fundamental process in a host of phenomena such as diffusion-controlled reactions~\cite{Smoluchowski1917,Kramers1940,Tachiya, Bramson, Szabo,Redner2001,Metzler2019,Isaacson2009, Benichou2010,Krapivsky2014}, binding of DNA transcription factors~\cite{Berg1981, Hippel2007, Loverdo2008, Benichou2011a}, animal foraging~\cite{Viswanathan1999}, the spread of infectious diseases, and fluctuations of stock prices~\cite{Bouchaud2003}.
	A central quantity characterizing the process is the first-passage time, namely, the time it takes a random walker or a diffusing particle to encounter a target for the first time.
	Over the past decades, the random-search problem has received considerable attention~\cite{Noh2004,Condamin2005,Condamin2007,Isaacson2013,Grebenkov2021,Ro2017},
	in particular, in the context of optimal search strategies~\cite{RoandKim2022,Bartumeus2002, Benichou2005, Evans2013, Tejedor2012, Jeanfrancois2016} and universal properties in scale-invariant processes~\cite{Condamin2007}; see, e.g., recent reviews~\cite{Benichou2014,Grebenkov2020} and references therein.

	While most studies rely on a single-particle picture, in practice, it is usual that a group of particles simultaneously search for a common target. In this case, the search time manifests the many-body properties even when searchers are noninteracting: The first-order statistic of the first-passage times recorded by individual searchers determines the search time and leads to the very characteristic dependence on the number of searchers~\cite{Weiss1983,Krapivsky1996,Mejia2011,Abad2012,Meerson2015,Ro2017,Grebenkov2020,LawleyDist,LawleyProb,LawleyUniv,RoandKim2022}. Moreover, searchers often interact with each other. One example is Mongolian gazelles utilizing acoustic communication to explore better habitat areas~\cite{Martinez2013}. For such interacting searchers, one of the fundamental questions is how inter-searcher interaction affects the first-passage time. Up to date, only a few studies tackle the interaction effect~\cite{Agranov2018,Park2020,Choi2021}.
	For example, Agranov and Meerson recently studied the narrow escape problem using the macroscopic fluctuation theory, where the presence of interaction is included implicitly through the diffusivity and mobility of the lattice gas particles~\cite{Agranov2018}.
	
	\begin{figure} [b]
		\center
		\includegraphics[width=1.0\linewidth]{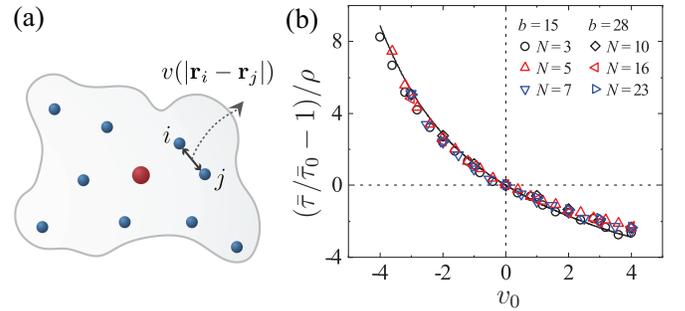}
		\caption{ (a) Schematic picture of Brownian particles (small blue circles) searching for a target (large red circle) in a finite domain.  Here, the particles interact with each other via a pairwise interaction $v(r)$. (b) Relative change of the global mean first-passage time, $\bar{\tau}/\bar{\tau}_0 -1$ vs. the interaction strength $v_0$ when $v(r) =k_\mathrm{B} T v_0   e^{-r/\sigma}$ with $\sigma = 0.5 a$. $\bar{\tau} (\bar{\tau}_0)$ is the search time averaged over a uniform initial searcher distribution in the presence (absence) of interactions, and $\rho \equiv (N-1)/V$. Langevin dynamics simulations are performed in two-dimensional circular domain of radius $b$ with a target size $a$ as a unit length, and each symbol indicates different $N$. Solid line represents the theoretical prediction given by Eq.~\eqref{Eq:1st_order} (see the main text).
		}
		\label{Fig:scheme}
	\end{figure} 
	
	To address the search problem in the presence of interactions in detail, we consider the dynamics of $N$ Brownian particles:
	\begin{equation}
		\label{eq:Langevin}
		 \mathbf{\dot r}_j = - \mu \nabla_{\mathbf{r}_j} U + \sqrt{2 D} \, {\boldsymbol \xi}_j (t) ,
	\end{equation}
	where $\mathbf{r}_j$ is the position of the $j$-th particle, $\mu$ is its mobility, $D=\mu k_B T$, and ${\boldsymbol \xi}_j (t)$ denotes the Gaussian white noise.
	The particles interact with each other via pairwise potential $v({\bf r})$, and the total interaction energy reads as $U({\cal R}) = \sum_{i<j}^N v({\bf r}_i - {\bf r}_j)$ with ${\cal R} = ({\bf r}_1, {\bf r}_2, \cdots, {\bf r}_N)$.
	We regard the target as being found when one of the searchers reaches the target.
	Direct simulations of Eq.~\eqref{eq:Langevin} for particles interacting via
	exponential potential show that the relative change of averaged search time depends on whether the interaction is attractive or repulsive (see Fig.~1).
	Then an imminent question arises: 
	Is it possible to analytically account for the effect of interaction if the interaction potential is given?
	This question becomes more intriguing when the interaction has both attractive and repulsive parts, which is the case for many colloidal particles.

	In what follows, we address these questions and provide a solution to the target search problem of interacting particles described by Eq.~\eqref{eq:Langevin}.
	To this end, we obtain the integral equation for the mean first-passage time (MFPT), which explicitly includes the interaction potentials. The potential is introduced in the form of the Mayer function which allows us to treat the potentials 
	even if they diverge due to hardcore repulsions at short distances.
	Then, solving the equation iteratively, we find that in the large volume limit, 
	the leading order correction to the MFPT arising from interactions is given in the form of virial expansion, and 
	does not depend on the other details of potentials and search domain shapes.
	
	\emph{Setting and formulation}---We have $N$ Brownian particles searching for a small-size target ${\cal T}$ in a $d$-dimensional finite domain ${\cal D}$ (see Fig.~\ref{Fig:scheme}).
	The stochastic trajectories of the Brownian particles that start from the initial positions ${\cal R}_0 = ({\bf r}_{0;1}, {\bf r}_{0;2}, \cdots, {\bf r}_{0;N})$ are characterized by the $N$-particle probability density function $P_N ({\cal R}, t| {\cal R}_0)$ of the particle positions ${\cal R}$ at time $t$, obeying the Fokker-Planck equation~\cite{Gardiner2009,Risken1996,Kampen2007}:
	\begin{equation} \label{Eq:FPE}
		\partial_t P_N ( {\cal R} , t| {\cal R}_0 ) = {\cal L}_\mathrm{FP} ({\cal R}) P_N ( {\cal R} , t| {\cal R}_0 )~
	\end{equation}
	with the time derivative $\partial_t = \partial/\partial t$ and 
	the forward Fokker-Planck operator 
	${\cal L}_\mathrm{FP} ({\cal R}) =  \beta D \nabla \cdot [  \nabla U({\cal R}) ] +  D \nabla^2$,
	where $\beta = 1/k_\mathrm{B} T$ and $\nabla = \sum_{i=1}^N \nabla_{{\bf r}_i}$. 
	An absorbing boundary condition on the target boundary $\partial {\cal T}$ and a reflecting boundary condition on the domain boundary $\partial {\cal D}$  are imposed as $P_N ({\cal R}, t|{\cal R}_0) = 0$ if  ${\bf r}_i \in \partial {\cal T}$ 
	and $ \hat{\bf n} \cdot {\bf j} ({\cal R}, t|{\cal R}_0) = 0$ if ${\bf r}_i \in \partial {\cal D}$, where $\hat{{\bf n}}$ is a unit vector normal to the domain boundary, and ${\bf j} ({\cal R}, t|{\cal R}_0) = -D\big( \beta \nabla U({\cal R}) +  \nabla\big) P_N ({\cal R}, t|{\cal R}_0)$ is the probability current. 
	
	In evaluating the MFPT, it is convenient to use the backward Fokker-Planck equation:
	\begin{equation} \label{Eq:backwardFPE}
		\partial_t P_N ({\cal R}, t|{\cal R}_0) = {\cal L}^\dagger_\mathrm{FP} ({\cal R}_0) P_N ({\cal R}, t|{\cal R}_0)
	\end{equation}
	with the adjoint Fokker-Planck operator ${\cal L}^\dagger_\mathrm{FP} ({\cal R}_0) = -\beta D \nabla_0 U({\cal R}_0) \cdot \nabla_0 + D \nabla_0^2$, where $\nabla_0$ is the del operator applied on ${\cal R}_0$.  Integrating Eq.~\eqref{Eq:backwardFPE} over ${\cal R}$ and $t$, we obtain an adjoint equation for the MFPT, $\tau ({\cal R}_0)$, for a given initial configuration ${\cal R}_0$~\cite{Gardiner2009,Risken1996,Kampen2007}:
	\begin{equation} \label{Eq:Ltau}
		{\cal L}^\dagger_\mathrm{FP} ({\cal R}_0) \tau ({\cal R}_0) = -1~,
	\end{equation}
	where 
	$\tau ({\cal R}_0) = \int^\infty_0 \, d t \int d {\cal R} ~ P_N({\cal R}, t| {\cal R}_0)$ defines the search time.
	Here, we used the fact that in a finite domain, the target is eventually found, i.e. $\lim_{t \rightarrow \infty}  \int d {\cal R} ~ P_N({\cal R}, t| {\cal R}_0) =0$.
	The boundary conditions for $\tau ({\cal R}_0)$ are given as follows:
	$\tau ({\cal R}_0)=0$ if ${\bf r}_{0,i} \in \partial {\cal T}$ and $\hat{{\bf n}} \cdot \nabla_0 \tau({\cal R}_0)=0$ if ${\bf r}_{0,i} \in \partial {\cal D} $~\cite{Gardiner2009}.
	
	To proceed, we introduce a $dN$-dimensional Green function $G({\cal R}|{\cal R}_0)$ of the Laplace operator satisfying $\nabla^2 G({\cal R}|{\cal R}_0) =\delta ({\cal R} - {\cal R}_0)$~\cite{Barton2005} and the boundary conditions, $G({\cal R}|{\cal R}_0) =0$ if ${\bf r}_{0;i} \in \partial {\cal T}$ and $ \hat{{\bf n}} \cdot \nabla_0 G({\cal R}|{\cal R}_0)=0$ if ${\bf r}_{0;i} \in \partial {\cal D}$.  
	If the particles do not interact with each other, $U({\cal R})=0$, Eq.~\eqref{Eq:Ltau} becomes $D \nabla^2_0 \tau({\cal R}_0) = -1$, and its solution ~\cite{Condamin2005}
	\begin{equation} \label{Eq:tau0}
		\tau_0 ({\cal R}_0) = - \frac{1}{D} \int d {\cal R}\,  G({\cal R} | {\cal R}_0) 
	\end{equation}
	determines the search time of non-interacting searchers.

	In the case of interacting searchers, we find that MFPT can be written using the Green function as
	\begin{eqnarray}\label{tau}
		\tau({\cal R}_{0})=\tau_{0}({\cal R}_{0})+\beta \int d{\cal R} \, G({\cal R}|{\cal R}_{0})\nabla U({\cal R})\cdot \nabla \tau({\cal R}). \quad
	\end{eqnarray}
	Defining $\tau({\cal R}_{0})-\tau_{0}({\cal R}_{0})\equiv \tau_{U}({\cal R}_{0})$, we rewrite $\tau_{U}$ in a manner akin to the virial expansion using $M({\cal R}) \equiv e^{-\beta U({\cal R})}-1=\prod_{i < j} (1+ f_{ij}) -1$ with the Mayer function $f_{ij} = e^{-\beta v (\bf{r}_i -\bf{r}_j)} - 1$:
	\begin{eqnarray} 
		\tau_U ({\cal R}_{0}) &=&-\beta \int d{\cal R}\, G({\cal R}|{\cal R}_{0}) M({\cal R})\nabla U({\cal R})\cdot \nabla \tau({\cal R}) \nonumber \\
		&&-\int d{\cal R}\, G({\cal R}|{\cal R}_{0}) \nabla M({\cal R})\cdot \nabla \tau({\cal R})~.
	\end{eqnarray} 
	Exploiting the relation $-\beta \nabla U({\cal R})\cdot \nabla \tau({\cal R}) = -\nabla^{2}\tau({\cal R}) -1/D$ and integrating by parts, we get
	\begin{eqnarray}
		\label{Eq:del_tau}
		\tau_{U}({\cal R}_{0})&=&\int d{\cal R}\, G({\cal R}|{\cal R}_{0}) M({\cal R}) \nabla^{2}\tau_{0} \nonumber \\
		&&+ \int d{\cal R}\, M({\cal R}) \nabla G({\cal R}|{\cal R}_{0})\cdot  \nabla \tau({\cal R}) 
	\end{eqnarray}
	and obtain the integral equation for MFPT as
	\begin{equation} \label{Eq:Mayer}
		\tau({\cal R}_0)  = \tau_0 ({\cal R}_0) + \int d {\cal R}\, M({\cal R}) {\cal G}({\cal R}|{\cal R}_0)~,
	\end{equation}
	where ${\cal G}({\cal R}|{\cal R}_0) = - G({\cal R}|{\cal R}_0)/D + \nabla G({\cal R}|{\cal R}_0) \cdot \nabla \tau({\cal R})$.
	
	Our basic strategy to solve the integral equation (\ref{Eq:Mayer}) is to expand the search time in the powers of $M$, whereby we let $\tau({\cal R}_{0})=\sum_{n=0}\tau^{(n)}({\cal R}_{0})$ with $\tau^{(n)}$ containing the $n$th power of $M$. 
	First, neglecting the inter-particle interactions, the zeroth-order approximation for the MFPT is $\tau^{(0)} ({\cal R}_0) = \tau_0 ({\cal R}_0)$.
	Replacing $\tau$ on the right hand side of Eq.~\eqref{Eq:Mayer} with $\tau^{(0)}$, we obtain 
	\begin{equation}\label{firstO}
		\tau^{(1)}({\cal R}_{0}) =\int d{\cal R}\, M({\cal R})  \nabla \cdot \Big( G({\cal R}|{\cal R}_{0})  \nabla \tau_{0} ({\cal R}) \Big)~.
	\end{equation}
	Continuing iteration as such, we find $\tau^{(n\geq 2)}$ as
	\begin{eqnarray}\label{nthO}
		\tau^{(n\geq 2)}({\cal R}_{0})
		&=&\int d{\cal R}_1\, {\cal O}_{1} \cdots \int d{\cal R}_{n}\,{\cal O}_{n} \tau_{0}({\cal R}_{n})~, 
	\end{eqnarray}
	where ${\cal O}_{i}=M({\cal R}_i) \nabla_i G({\cal R}_i |{\cal R}_{i-1})  \cdot \nabla_i $. Summing up, the modifications of the search time caused by inter-searcher interactions rest on Eqs.~(\ref{firstO}) and (\ref{nthO}), and one can truncate the series upon the degree of pursued accuracy.

	\emph{Dilute Limit}---For dilute systems of interest in our study, the Mayer function $f ({\bf r})$ is negligible when $|{\bf r}|$ is longer than an interaction range (assumed short here) and serves as an expansion parameter. Collecting terms linear in $f$, which only appear in $\tau^{(1)}({\cal R})$, we obtain
	\begin{equation} \label{Eq:Mayer1st}
		\tau({\cal R}_0) = \tau_0({\cal R}_0) + \int d {\cal R}\, \sum_{i<j} f_{ij} \nabla \cdot \left[ G({\cal R}_0|{\cal R}) \nabla \tau_0 ({\cal R}) \right]. 
	\end{equation}
	This is the first main result.
	We note that we also obtain the same equation in an alternative approach;
	performing an expansion in terms of the interaction strength, we derive a perturbation equation of the search time as a series of Poisson equations, with an analogy to the electrostatic problems.
	Then, rearranging terms according to the volume order and gathering the leading order terms only, we arrive at the same equation (see details in SI~\cite{supp}).

	In cases where particles uniformly populate the searching domain at the initial time, the average search time over a uniform initial distribution, called the global mean first-passage time (GMFPT), is given by
	\begin{eqnarray} \label{Eq:GMFPT}
		\bar{\tau} \equiv \frac{1}{V^N} \int d {\cal R}_0\, \tau({\cal R}_0) .
	\end{eqnarray}
	Here, for convenience, we take the target volume as a unit and the volume of ${\cal D}$ as $V$. 
	Inserting Eq.~\eqref{Eq:Mayer1st} into Eq.~\eqref{Eq:GMFPT}, we obtain the GMFPT for interacting particles:
	\begin{equation} \label{Eq:taubar}
		\bar{\tau} (v_0) \simeq \bar{\tau}_0 - \frac{D}{V^N} \int d {\cal R}\, \sum_{i<j}^N f_{i j} \nabla \cdot \left[ \tau_0({\cal R}) \nabla \tau_0 ({\cal R}) \right] ~. 
	\end{equation}

	To tackle the analytic calculation of Eq.~(\ref{Eq:taubar}), an instrumental element is $
	\tau_{0}({\cal R})$ determined by the Green function $G({\cal R}^{\prime}|{\cal R})$ through the relation (\ref{Eq:tau0}). The exact expression of $G({\cal R}^{\prime}|{\cal R})$ reads as
	\begin{equation} \nonumber
		G({\cal R}^{\prime} |{\cal R}) = \sum_{n_1, \dots, n_N=1}^\infty \left(\frac{-1}{\sum_{i=1}^N \lambda_{n_i} }\right) \prod_{i=1}^N \Big( \psi_{n_i} ({\bf r}^{\prime}_i) \psi_{n_i} ({\bf r}_{i}) \Big)~,
	\end{equation}
	where $\psi_k ({\bf r})$ and $\lambda_k$ are $k$-th eigenfunction and eigenvalue of the Laplace operator, respectively~\cite{comment}
	The eigenvalues are arranged in ascending order in magnitude.
	Noting that in the limit of $V\rightarrow \infty$ the eigenvalue $\lambda_{1}$ with the smallest magnitude  approaches zero, and other eigenvalues remain finite~\cite{Pinsky2003}, we can simplify the above expression considerably for the case of our interest where the search domain is much larger than the target size:

	\begin{eqnarray}\label{Eq:G_largeV} 
		G({\cal R}^{\prime}|{\cal R}) \simeq  -\frac{1}{N \lambda_1} \prod_{i}\psi_1({\bf r}^{\prime}_i) \psi_1 ({\bf r}_{i})~.
	\end{eqnarray}
	This asymptotic expression of the Green function leads to the search time of non-interacting searchers as
	\begin{equation}\label{asymptau0}
		\tau_{0}({\cal R})
		\approx \frac{V^{N/2}}{ND\lambda_{1}}\prod_{i}^{N} \psi_{1}({\mathbf r}_{i})~,
	\end{equation}
	where we let $\int d{\mathbf r}\, \psi_{1}({\bf r}) \simeq V^{1/2}$, using the fact that the normalized eigenfunction $\psi_{1}$ is almost constant except the region close to the target.
	Integrating $\tau_{0}({\cal R})$ over ${\cal R}$, one from Eq.~(\ref{Eq:GMFPT}) obtains the well-known expression of GMFPT of non-interacting searchers, ${\bar \tau}_{0}\simeq (N\lambda_{1}D)^{-1}$~\cite{Ro2017}.

	Putting Eq.~(\ref{asymptau0}) into Eq.~(\ref{Eq:taubar}) and performing the integrations allows us to estimate the interaction effects. 
	In the case of a dilute system with an interaction  range relatively shorter than the system size, by approximating the Mayer function as a delta function and following steps detailed in SI~\cite{supp}, we obtain the second main result, an expression for GMFPT:
	\begin{equation} \label{Eq:1st_order}
		\bar{\tau}  = \bar{\tau}_0 \left[ 1 
		- \frac{4}{3}  \frac{N-1}{V} B_2 \right]~,
	\end{equation}
	where $B_{2}$ is the second virial coefficient 
	\begin{equation}
		B_{2}=-\frac{1}{2}\int d^{d}{\bf r}\, [e^{-\beta v(|{\bf r}|)}-1]~. 
	\end{equation}
	Note that the first correction term on the search time due to interaction is proportional to the particle density and the second virial coefficient of the interaction potential, and does not depend on other details such as domain shapes.
	The relation \eqref{Eq:1st_order} predicts how the sign of interaction affects the search time. 
	For repulsive interactions, the second virial coefficient is positive, and the search time is shorter than the non-interacting, ideal case and decreases with elevated interaction strengths. Attractive interactions act oppositely and slow down the search process. 
	This is consistent with our numerical results presented in Fig.~1 and can be intuitively understood by considering how the particle distribution close to the target boundary changes as the interaction turns on. In terms of searcher distribution, the target area is a cavity devoid of particles, causing an imbalance in force distributions. Therefore, particles repelling one another are pushed toward the target, shortening MFPT, whereas attracting particles get shoved away from the target and take longer MFPT.
	
	\begin{figure} [t]
		\center
		\includegraphics[width=0.8\linewidth]{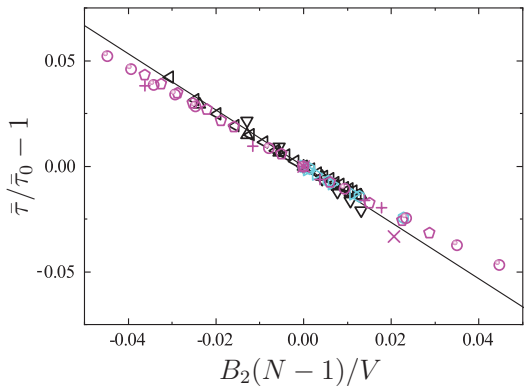}
		\caption{Relative change of the global mean first-passage time vs. the second virial coefficient ($B_\mathrm{2}$) times the searcher density.
			The global mean first-passage times (symbols) are measured through the Langevin dynamics simulations for three different types of inter-searcher potentials;
			i) exponential (black), $v(r)= k_\mathrm{B} T v_0 e^{-r/\sigma}$, ii) hardcore (cyon) of a diameter $\sigma$, iii) Sutherland (magenta), i.e., hardcore for $r<\sigma$
			and $-k_B T v_0 (\sigma/r)^6$, otherwise.
			The simulations are performed for a broad range of parameters, e.g., various numbers of particles ($N=3 \sim 23$) and interaction parameters ($v_0$ and $\sigma$) are used. 
			Solid line represents our theoretical prediction given by Eq.~\eqref{Eq:1st_order}.
			In simulations, a two-dimensional circular domain of radius $b= 15a \sim 35a$ (volume $V$) with a small circular target of radius $a$ at the center is considered,
			and the statistical average is taken over more than $10^7$ ensembles to obtain each point. 
		}
		\label{Fig:time}
	\end{figure}

	\emph{Simulations and Discussion}---To verify our prediction~Eq.~\eqref{Eq:1st_order} quantitatively, we perform the overdamped Langevin dynamics simulations, considering $N$ particles with a uniform initial distribution in a domain with a small target located in it, and measure GMFPT. 	
	In simulations, we consider three different potentials; i) exponential, $v({\bf r}) = v_0 k_\mathrm{B} T \exp(-|{\bf r}|/\sigma)$ with an interaction strength $v_0$ and a decay length $\sigma$, ii) hardcore potential of radius $\sigma$, $v ({\bf r}) = \infty$ for $|{\bf r}|  < \sigma$ and $0$, otherwise, and
	iii) Sutherland potential, $v ({\bf r}) = \infty$ for $|{\bf r}|  < \sigma$ and $- v_0 k_\mathrm{B} T  \left( \sigma/r \right)^6$, otherwise. 
	In simulatting hardcore interactions, we let the particles repel along the line connecting their centers until there is no overlap between them. 
	To demonstrate the universal feature, we plot the relative changes in GMFPTs together in Fig.~2 for all different potentials.
	For various values of the searcher number $N$, the domain size $b$, the interaction range $\sigma$, and the interaction strength $v_0$ considered here,
	all data points show a good agreement with the theoretical prediction, Eq.~\eqref{Eq:1st_order}, represented as a solid line, clearly demonstrating the validity of our theory.
	Formulated in terms of the Mayer function, our theory applies to diverging potentials at short distances, such as the hardcore potential.
	Intermolecular potentials usually consist of a short-range hardcore repulsion and a smooth, relatively long-range attraction, as in DLVO theory.
	For the potential of such type combining repulsive and attractive interactions, it would not be obvious, even at a qualitative level, to predict the effect of interactions on the search time.
	Our result of Eq.~\eqref{Eq:1st_order}, however, provides an explicit answer; the second virial coefficient determines the search time.
	
	Our main result considers the search time in the large volume limit, or equivalently, the dilute density regime, which is the most relevant in the context of the target search. In semi-dilute regimes, the higher-order terms neglected in Eq.~\eqref{Eq:taubar} would contribute, and the GMFPT deviates from the linear, leading order behavior. For example, in the case of strongly attracting particles, the aggregation occurs as the particle density increases, and then the searching dynamics may show abrupt changes as a result of the clustering of the particles, as reported in Ref.~\cite{Choi2021}.
	In deriving Eq.~\eqref{Eq:1st_order}, we also assumed the short-ranged potential. Otherwise, the integral in evaluating $B_\mathrm{2}$ is dominated by large distances and may diverge. Therefore, in the case of long-range interactions such as Coulomb potential, the search time is inexpressible in the form of a virial series.

	We remark that for soft potentials, we may expand the Mayer function as $\exp[-\beta v({\bf r})] - 1 \simeq -\beta v({\bf r})$, and the second virial coefficient of Eq.~\eqref{Eq:1st_order} is replaced by the integral of the potential. 
	In fact, adopting the action derived from the Dean's equation~\cite{Dean1996} into the large deviation functional considered in Ref.~\cite{Agranov2018}, we can obtain the equation similar to Eq.~\eqref{Eq:1st_order}, for soft potentials.
	In such an approach, however, the virial coefficient does not appear explicitly in the expression of searching time.
	
	\emph{Summary}---We have studied random target searches by interacting Brownian particles in a confined space. 
	As the target search usually becomes important when the searcher density is low, we considered expansion using the Mayer function and derived the leading order correction in the dilute searcher density limit. 
	For short-ranged interactions where the Mayer function is approximated as a delta function, we have derived the leading order correction on the GMFPT which is proportional to the second virial coefficient of the potential and the particle density. 
	Lastly, we have verified our theory by comparing the theoretical expectations with the Langevin dynamics simulations. 
	In future studies, it would be interesting to extend the formalism suggested in the paper to consider the random target searching of particles with long-range interaction and Kramers barrier crossing problem for interacting particles. 
	
	We thank M. Kardar for helpful discussions and a careful reading of the manuscript.
	This research was supported by a National Research Foundation of Korea (NRF) grant funded by the Korean government (Grant No. NRF-2020R1A2C1014826).


\begin{thebibliography}{100} 

\bibitem{Smoluchowski1917} M. V. von Smoluchowski, Z. Phys. Chem. \textbf{92}, 129 (1917).

\bibitem{Kramers1940} H. A. Kramers, Physica \textbf{7}, 284 (1940).

\bibitem{Tachiya}
M. Tachiya, Radiat. Phys. Chem. {\bf 21}, 167 (1983).

\bibitem{Bramson} 
M. Bramson and J. L. Lebowitz, Phys. Rev. Lett. {\bf 61}, 2397 (1988).

\bibitem{Szabo}
A. Szabo, R. Zwanzig, and N. Agmon, Phys. Rev. Lett. {\bf 61}, 2496 (1988).

\bibitem{Redner2001}
S. Redner, \emph{A Guide to First Passage Processes} (Cambridge University Press, Cambridge, 2001).

\bibitem{Krapivsky2014}
B.~Meerson, A.~Vilenkin, and P.~L.~Krapivsky, Phys. Rev. E {\bf 90}, 022120 (2014).

\bibitem{Benichou2010}
O. B\'{e}nichou, C. Chevalier, J. Klafter, B. Meyer, and R. Voituriez, Nat. Chem. {\bf 2}, 472 (2010). 

\bibitem{Metzler2019}
R. Metzler, J. Stat. Mech. 114003 (2019).

\bibitem{Isaacson2009}
S.~A.~Isaacson and D.~Isaacson, Phys. Rev. E \textbf{80}, 066106 (2009).

\bibitem{Berg1981} 
O. G. Berg, R. B. Winter, and P. H. von Hippel, Biochemistry \textbf{20}, 6929 (1981).

\bibitem{Hippel2007} P. H. von Hippel, Annu. Rev. Biophys. Biomol. Struct. \textbf{36}, 79 (2007).

\bibitem{Benichou2011a}
O. B\'{e}nichou, C. Chevalier, B. Meyer, and R. Voituriez, Phys. Rev. Lett. \textbf{106}, 038102 (2011).

\bibitem{Loverdo2008}
C. Loverdo, O. B\'{e}nichou, M. Moreau, and R. Voituriez, Nat. Phys. {\bf 4}, 134 (2008). 


\bibitem{Viswanathan1999} G. M. Viswanathan, S. V. Buldyrev, S. Havlin, M. G. E. Da Luz, E. P. Raposo,  and H. E. Stanley, Nature (London) \textbf{401}, 911 (1999).

\bibitem{Bouchaud2003} J. P. Bouchaud, M. Potters, \textit{Theory of Financial Risk and Derivative Pricing: From Statistical Physics to Risk Management} (Cambridge University Press, Cambridge, UK, 2003).





\bibitem{Noh2004} J. D. Noh and H. Rieger, Phys. Rev. Lett. \textbf{92}, 118701 (2004).

\bibitem{Condamin2005}
S. Condamin,  O. B\'{e}nichou, and M. Moreau, Phys. Rev. Lett. \textbf{95}, 260601 (2005).

\bibitem{Condamin2007}
S. Condamin, O. B\'{e}nichou, V. Tejedor, R. Voituriez, and J. Klafter, Nature \textbf{450}, 77 (2007).

\bibitem{Isaacson2013}
S.~A.~Isaacson and J.~Newby, Phys. Rev. E \textbf{88}, 012820 (2013).

\bibitem{Grebenkov2021}
D. S. Grebenkov, R. Metzler, and G. Oshanin, New J. Phys. \textbf{23}, 123049 (2021). 


\bibitem{Ro2017} 
S. Ro and Y. W. Kim, Phys. Rev. E \textbf{96}, 012143 (2017).

\bibitem{RoandKim2022} S. Ro and Y. W. Kim, Phys. Rev. E \textbf{106}, 024101 (2022).

\bibitem{Bartumeus2002} F. Bartumeus, J. Catalan, U. L. Fulco, M. L. Lyra, and G. M. Viswanathan, Phys. Rev. Lett. \textbf{88}, 097901 (2002).

\bibitem{Benichou2005} O. B\'{e}nichou, M. Coppey, M. Moreau, P-H. Suet, and R. Voituriez, Phys. Rev. Lett. \textbf{94}, 198101 (2005).

\bibitem{Evans2013} M. R. Evans, S. N. Majumdar, K. Mallick, J. Phys. A: Math. Theor. \textbf{46}, 185001 (2013).

\bibitem{Tejedor2012} V. Tejedor, R. Voituriez, O. B\'enichou, Phys. Rev. Lett. \textbf{108} (2012).

\bibitem{Jeanfrancois2016} J. Rupprecht, O. B\'enichou, R. Voituriez, Phys. Rev. E \textbf{94}, 012117 (2016).

\bibitem{Benichou2014} O. B\'{e}nichou and R. Voituriez, Phys. Rep. {\bf 539}, 225 (2014).

\bibitem{Grebenkov2020} D. S. Grebenkov, R. Metzler, and G. Oshanin, New J. Phys. \textbf{22}, 103004 (2020). 

\bibitem{Weiss1983}
G. H. Weiss, K. E. Shuler, and K. Lindenberg, J. Stat. Phys. {\bf 31}, 255 (1983).

\bibitem{Krapivsky1996}
P. L. Krapivsky and S. Redner, J. Phys. A {\bf 29}, 5347 (1996).

\bibitem{Mejia2011}
C. Mejía-Monasterio, G. Oshanin, and G. Schehr, J. Stat. Mech. Theory Exp. P06022 (2011).

\bibitem{Abad2012}
E. Abad, S. B. Yuste, K. Lindenberg, Phys. Rev. E \textbf{86}, 061120 (2012).

\bibitem{Meerson2015}
B. Meerson and S. Redner, Phys. Rev. Lett. \textbf{114}, 198101 (2015).

\bibitem{LawleyDist}
S. D. Lawley, J. Math. Biol. \textbf{80}, 2301 (2020).

\bibitem{LawleyProb}
S. D. Lawley and J. B. Madrid, J. Nonlinear Sci. \textbf{30}, 1207 (2020).

\bibitem{LawleyUniv}
S. D. Lawley, Phys. Rev. E \textbf{101}, 012413 (2020). 

\bibitem{Martinez2013}
R. Martínez-García, J. M. Calabrese, T. Mueller, K. A. Olson, and C. López, Phys. Rev. Lett. \textbf{110}, 248106 (2013).

\bibitem{Agranov2018}
T. Agranov and B. Meerson, Phys. Rev. Lett. {\bf 120}, 120601 (2018).

\bibitem{Park2020}
S.-C. Park, Phys. Rev. E {\bf 102}, 042112 (2020).

\bibitem{Choi2021}
M. Choi, Y. W. Kim, J. Korean Phys. Soc. {\bf 79}, 653 (2021).

\bibitem{Gardiner2009}
C. W. Gardiner, \emph{Stochastic Methods}, 4th ed. (Springer-Verlag, Berlin, Heidelberg, New York, 2009). 

\bibitem{Kampen2007}
N. G. Van Kampen, \emph{Stochastic Processes in Physics and Chemistry}, 3rd ed. (North-Holland, Amsterdam, 2007).

\bibitem{Risken1996}
H. Risken, \emph{The Fokker-Planck Equation}, 2nd ed. (Springer, New York, 1996). 

\bibitem{Barton2005}
G. Barton, \emph{Elements of Green's functions and propagation: potentials, diffusion, and waves}, (Oxford University Press, Oxford, 1989).

\bibitem{Pinsky2003} R. Pinsky, J. Funct. Anal. \textbf{200}, 177 (2003).

\bibitem{supp} See Supplemental Material [url], which includes theoretical and numerical details.

\bibitem{comment} The eigenfunctions satisfy the absorbing boundary condition $\psi_k({\bf r}) = 0$ at ${\bf r} \in \partial {\cal T}$ and the reflecting boundary condition $\hat{\bf n} \cdot \nabla_{\bf r} \psi_k ({\bf r}) = 0$ at ${\bf r} \in \partial {\cal D}$, complying with the boundary conditions imposed on the Green function.

\bibitem{Dean1996}
D. S. Dean, J. Phys. A Math. Gen. \textbf{29}, L613 (1996).

\end{thebibliography}
\end{document}